# STC Anti-spoofing Systems for the ASVspoof 2015 Challenge

*Sergey Novoselov* [1,2], *Alexandr Kozlov*[2], *Galina Lavrentyeva*[2], *Konstantin Simonchik* [1,2], *Vadim Shchemelinin*[1,2]

[1]ITMO University, St. Petersburg, Russia
[2]Speech Technology Center Ltd., St. Petersburg, Russia

`{novoselov, kozlov-a, lavrentyeva, simonchik, shchemelinin}@speechpro.com`

## Abstract

This paper presents the Speech Technology Center (STC) systems submitted to Automatic Speaker Verification Spoofing and Countermeasures (ASVspoof) Challenge 2015. In this work we investigate different acoustic feature spaces to determine reliable and robust countermeasures against spoofing attacks. In addition to the commonly used front-end MFCC features we explored features derived from phase spectrum and features based on applying the multiresolution wavelet transform. Similar to state-of-the-art ASV systems, we used the standard TV-JFA approach for probability modelling in spoofing detection systems. Experiments performed on the development and evaluation datasets of the Challenge demonstrate that the use of phase-related and wavelet-based features provides a substantial input into the efficiency of the resulting STC systems. In our research we also focused on the comparison of the linear (SVM) and nonlinear (DBN) classifiers.

## 1. Introduction

Information technology plays an increasingly large role in today's world, and different authentication methods, including voice biometrics, are used for restricting access to informational resources. Examples of speaker recognition systems usage include internet banking systems, customer identification during a call to a call center, as well as passive identification of a possible criminal using a preset "blacklist" [1], [2]. Due to the importance of the information that needs to be protected, requirements for biometric systems are high, including robustness against potential break-ins and other attacks.

Performance of basic technologies in voice biometrics has greatly improved in recent years. For instance, the latest overviews of speaker recognition systems showed that EER is down to 1.5-2% for text-independent [3] and down to 1% for text-dependent [4] speaker recognition systems in various conditions.

With the growth of interest in reliable ASV systems, the development of their spoofing techniques increased tremendously [1]. A multitude of different spoofing methods was proposed in literature. For example, [5] describes methods based on "Replay attack", "Cut and paste", "Handkerchief tampering" and "Nasalization tampering". Speech synthesis approaches [6] are also widely used for spoofing purposes.

Despite the development of new spoofing detection methods, most of ASV spoofing countermeasures presented so far are dependent on a training dataset related to a specific type of attack, while the nature of the spoofing attack is usually unknown in real practice. Several papers on the ASV system robustness evaluation against spoofing attacks [5], [7], [8] show that it is highly important to develop new anti-spoofing techniques to detect unforeseen spoofing attacks when details of the spoofing attacks are unknown, in order to keep the required EER level. That was the motivation for organizing the ASVspoof Challenge 2015 [9] where spoofing detection methods for known [1] and unknown spoofing types were evaluated. ASVspoof Challenge 2015 was focused on a stand-alone spoofing detection task.

In this paper we describe several spoofing detection systems that were proposed for the ASVspoof Challenge 2015.

For participation in the challenge we used the standard Total Variability Joint Factor Analysis (TV-JFA) approach for statistical modelling of the acoustic features of the speech signal. As a classifier we applied Support Vector Machine (SVM) or, alternatively, Deep Belief Network (DBN).

In the paper we concentrate on researching the most appropriate front-end features for the spoofing detection system we propose. In particular, we investigated acoustic features based on the phase spectrum information and features derived by applying the wavelet transform [10]. The aim of our research was to find the most effective method for detecting unknown spoofing attacks.

The remainder of this paper is organized as follows. Section 2 overviews all the proposed spoofing detection systems with a brief description of the subsystems it consists of. Section 3 introduces acoustic feature extraction methods in detail. Experimental work is described in Section 4. Finally, our conclusions are presented in Section 5.

## 2. Overall system description

All our systems consist of three main components (Figure 1):
- Acoustic feature extractor
- TV i-vector extractor
- Classifier

After experiments on the training part of the ASVspoof Challenge 2015 database we decided to include pre-detection as a preliminary step in the spoofing detection system. The pre-detector checks whether the input speech signal has zero

This work was partially financially supported by the Government of Russian Federation, Grant 074-U01.

temporal energy values. In case of zero-sequence the signal is declared to be a spoofing attack, otherwise (or in case of a system without the pre-detection step) the speech signal is used as input data for the feature extractor.

Acoustic feature extractors in our systems combine several different acoustic feature extraction methods. All the features we proposed during the ASVspoof Challenge 2015 will be described in details in Section 3.

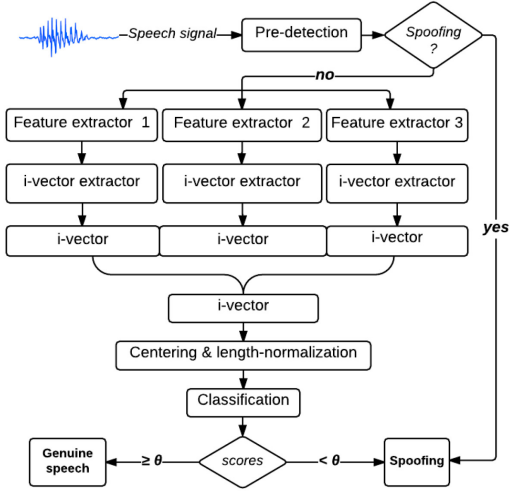

Figure 1: *Scheme of spoofing detection system.*

The obtained feature vectors are used by the i-vector extractor to get i-vectors from different feature types. These i-vectors are then concatenated in one common i-vector, which is centered and length-normalized.

Finally, the classifier calculates the resulting score to estimate if the speech signal is a spoofing attack or not (Figure 1).

### 2.1. Total Variability Modeling

In our work for the acoustic space modelling we used the standard TV-JFA approach, which is the state-of-the-art in speaker verification systems [3], [4], [11]. According to this version of the joint factor anlysis, the i-vector of the Total Variability space is extracted by means of JFA modification, which is a usual Gaussian factor analyser defined on mean supervectors of the Universal Background Model (UBM) and Total-variability matrix T. UBM is represented by the Gaussian mixture model of the described features. The diagonal covariance UBM was trained by the standard EM-algorithm.

### 2.2. SVM-Classification

In our work we used an SVM classifier with a linear kernel. The separating hyperplane was trained in normalized i-vectors space (Figure 1) on the training part of the ASV spoof challenge 2015 database to detect genuine speech phonograms and spoofing attacks. In SVM training the efficient LIBLINEAR [12] library was used for calculations in order to achieve the necessary accuracy and computational speed.

### 2.3. DBN-Classification

Alternatively, we used a classifier based on Deep Belief Network with softmax output units and stochastic binary hidden units [13] (Figure 2). DBN takes normalized i-vectors from the i-vector extractor as input data. We used layer-wise pretreating of the layers by means of Restricted Boltzmann Machines (RBMs) and then applied back-propagation to train the DBN in a supervised way to perform classification.

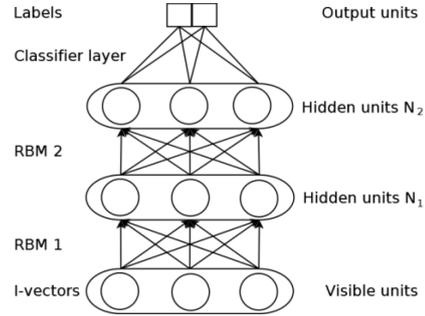

Figure 2: *DBN-Classification.*

## 3. Front-End Features

In this section we describe a number of different acoustic features that were effective for the ASVspoof Challenge 2015 task.

### 3.1. Amplitude spectral features

As short-term amplitude spectrum acoustic features to be used in the ASV spoofing attack detection system we selected mel frequency coefficients. We used the discrete cosine transform to obtain Mel-Frequency Cepstral Coefficients (MFCC) and principal component analysis to obtain Mel-Frequency Principal Coefficients (MFPC). These features accurately represent the general characteristics of the vocal tract.

MFCC coefficients represent the short-term power spectrum of the speech signal, based on the application of the discrete cosine transform to the log power spectrum on a nonlinear mel scale of frequency (Figure 3). To derive the MFCC coefficients we used a Hamming window function with the window length equal to 256 and 50% overlap.

We kept the first 12 MFCC coefficients and their first and second-order derivatives as the most informative acoustic features, thereby obtaining a 36-length feature vector.

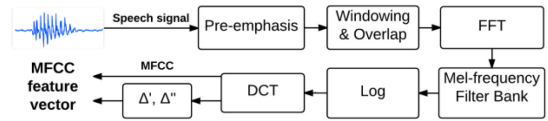

Figure 3: *MFCC feature extractor.*

The MFPC coefficients were obtained similarly to the MFCC coefficients, but using principal component analysis instead of the discrete cosine transform to achieve decorrelation of the informative acoustic features (Figure 4).

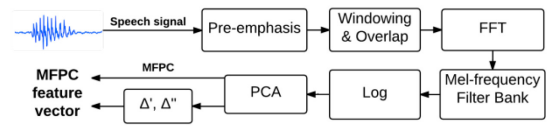

Figure 4: *MFPC feature extractor.*

### 3.2. Phase-based features

In order to take into account phase information of the speech signal we used the cos-phase features described in [14]. These features were extracted from the phase spectrum, obtained by the Fourier Transform, as follows:

1. The unwrapped phase spectrum was normalized by applying the cosine function to change its range to [-1; 1].
2. Dimensionality reduction was then performed by means of principal component analysis, the basis of which had been calculated beforehand on the training part of the ASVspoof Challenge 2015 database.

Similarly to amplitude spectrum features, we selected only the first 12 coefficients with their first and second-order derivatives to form the resulting feature vector CosPhase Principal Coeffitients (CosPhasePC).

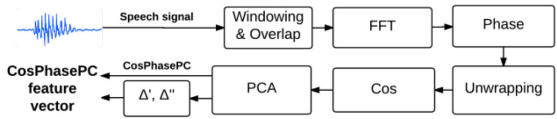

Figure 5: *CosPhasePC feature extractor.*

### 3.3. Wavelet-based features

We decided to introduce detailed time-frequency analysis of the speech signals in our countermeasures. For this purpose we used front-end features based on applying the wavelet-packet transform [10], adapted to the mel scale (Figure 6).

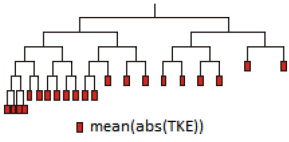 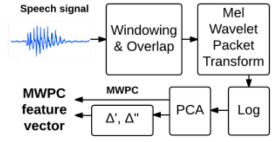

Figure 6: *Wavelet Packet Transform. TKE is a Teager Kaiser energy*

Figure 7: *MWPC feature extractor*

Instead of the classical energy of the frequency sub-bands, here we applied the Teager Keiser Energy (TKE) Operator (Figure 6). TKE is more informative than classical sample energy. Moreover, it is a noise-robust parameter for speech signal [15]. We used the following equation for TKE evaluation:

$$\Psi(s(t)) = s(t)^2 - s(t-1)s(t+1) \quad (1)$$

where $s(t)$ is the output signal temporal sample of the considered sub-band.

For extracted features decorrelation, similarly to 3.1 and 3.2, we consistently applied principal component analysis to derive 12 coefficients. We called these features Mel Wavelet Packet Coefficients (MWPC) for short. Equivalently to 3.1 and 3.2, here we observe MWPC with its first and second-order derivatives (Figure 7).

To derive the MWPC coefficients we also used a Hamming window function with the window length equal to 256 and 50% overlap.

## 4. Experiments

For training all parameters of our anti-spoofing systems we used training dataset. All the experiments, described below, were performed on the development dataset. According to the challenge conditions [9], the training and development datasets contained 5 spoofing attacks of types S1-S5: S1,S2,S5 were voice conversion algorithms and S3, S4 were HMM-based speaker-adapted speech synthesis methods.

For example, Figure 8 presents LDA projections of the MWPC based i-vectors for the subset of the development dataset on the 3 principal components $P_1$, $P_2$, $P_3$, evaluated on the training dataset. In our experiments we used Daubechies wavelets db4 for wavelet transform.

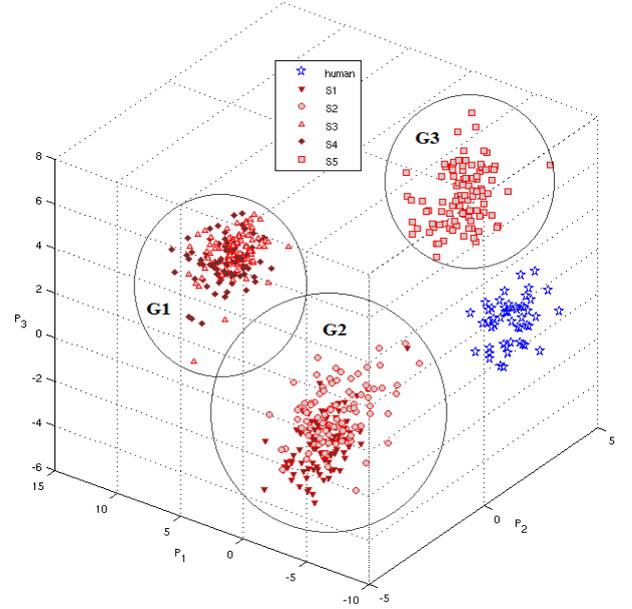

Figure 8: *The LDA projections of MWPC based i-vectors on principal components $P_1$, $P_2$, $P_3$.*

In Figure 8 we can see that that human class is well separated from the spoofing classes. This allows us to argue that integration of SVM linear classification methods can be efficient for solving the spoofing detection task in this space. Note that in this feature space 3 groups of spoofing classes can also be easily discriminated: G1 is the group of HMM-based speech synthesis (S3+S4); G2 is the group of simple Voice conversion techniques (S1+S2); G3 represents Festvox voice conversion method.

Table 1 demonstrates resulting EER estimates (%) of the TV-SVM based system with different front-end features, described in Section 3, obtained on the development dataset. These results were obtained with the use of 256-component UBM and 200 dimensional i-vector.

Table 1. *Experimental results for the TV-SVM systems for different types of spoofing algorithms, EER (%).*

| Features | Spoofing algorithm | | | | | |
| --- | --- | --- | --- | --- | --- | --- |
| | *S1* | *S2* | *S3* | *S4* | *S5* | *All* |
| MFCC | 0.38 | 2.13 | 0.36 | 0.39 | 1.48 | 1.14 |
| MFPC | 0.13 | 0.29 | 0.09 | 0.09 | 0.37 | 0.23 |
| CosPhasePC | 0.13 | 0.20 | 0.04 | 0.05 | 0.23 | 0.15 |
| **MWPC** | **0.03** | **0.11** | **0.00** | **0.00** | **0.08** | **0.05** |

The results show that standard MFCC features are inferior in comparison with other front-end features by EER estimates. It should be mentioned that substantial EER improvement was achieved on MFPC for all spoofing techniques by PCA basis applying in MFCC instead of discrete cosine transform. CosPhasePC implementation slightly reduces EER compared with MFPC. Features based on the multiresolution wavelet transform outperform all other proposed features, reaching 0.05% EER for all known attacks. Note that during our experiments we determined that using the TKE operator in MWPC demonstrates better results than classical energy.

### 4.1. Fusion

In this work we explored fusion of systems based on different front-end features in the i-vector space. All i-vectors derived from different extractors were concatenated in one common i-vector, as shown in Figure 1. Results for different fusion systems confirm that for all combinations of features EER estimations for known spoofing attacks are close to 0%. However, the zero error of spoofing detection for the development dataset was achieved only by implementing the combination of CosPhasePC and MWPC features.

### 4.2. Classifiers

Experiments of the fused TV systems with the combination of MFCC, MFPC and CosPhasePC feature extractors based on SVM and DBN classifiers showed that SVM-based system achieved 0.03% EER on the development dataset, which is better than 0.04% EER of DBN-based system.

## 5. Evaluation results

Based on the results of our experiments on the development dataset, we decided to propose 3 systems according to the common condition of the ASVspoof Challenge 2015.

Our **Primary** system was implemented according to the scheme on Figure 1 with a pre-detection step and MFCC, MFPC and CosPhasePC as acoustic features. The UBM was represented by a 1024-component Gaussian mixture model of the described features, and the dimension of the TV space was 400. Here we used SVM for classification.

In our **Contrastive 1** system a pre-detector was not used, and MWPC features were used instead of MFCC.

Our third system **Contrastive 2** also did not use a pre-detector and applied a nonlinear DBN classifier. In order to avoid overfitting, we reduced the UBM component number to 256 for all feature types and the TV space dimension to 200.

In addition to known spoofing attacks, the evaluation dataset contained spoofing attacks of 5 unknown types S6-S10 [16]. Table 2 presents the evaluation results.

Table 2. *Evaluation results of the STC systems, EER (%).*

| System | Known attacks | Unknown attacks | All |
|---|---|---|---|
| Primary | **0.008** | **3.922** | **1.965** |
| Contrastive 1 | 0.009 | 4.891 | 2.450 |
| Contrastive 2 | 0.017 | 6.162 | 3.090 |
| Primary (no pre-detection) | 0.008 | 5.151 | 2.579 |

In spite of the good results of the STC systems for known attacks, results obtained for unknown attacks are much worse: our best primary system reached 3.92% EER. This observation suggests the necessity to improve countermeasures to achieve robust performance on unknown attacks.

The primary system showed the best results, in particular due to the energy pre-detection step. To confirm this suggestion we compared this result with EER for primary system without pre-detector (Table 2). This pre-detector will not be useful in case of channel effects or additive noise. That is a significant limitation of this anti-spoofing system.

Unlike the Primary system, the Contrastive 1 system did not use the pre-detector and demonstrated relatively good performance according to the challenge results. We see the reason for that in the MWPC features. The advantage of these features is the wavelet-transform, which makes it possible to produce detailed multiresolution signal analysis. That provides an additional decrease of EER in the spoofing detection task.

Results for the Contrastive 2 system with DBN classifier, turned out to be the worst. In this system we probably failed to avoid the effects of the stronger overfitting on these training dataset, in comparison with SVM. According to the results it can be suggested that it is better to use linear SVM classifier in the proposed system (Figure 1).

In Table 3 the evaluation EER results for each spoofing type are introduced. It can be seen that results of all proposed systems for unknown spoofing types S6-S9 and for known types S1-S5 are comparable. But for S10() EER results of all our systems are several orders greater. That affects dramatically the overall EER evaluation results.

Table 3. *Evaluation results for STC systems for different types of spoofing algorithms, EER (%).*

| Spoofing algorithm | System | | |
|---|---|---|---|
| | Primary | Contrastive 1 | Contrastive 2 |
| S1 | 0.004 | 0.005 | 0.000 |
| S2 | 0.022 | 0.022 | 0.058 |
| S3 | 0.000 | 0.000 | 0.000 |
| S4 | 0.000 | 0.000 | 0.000 |
| S5 | 0.013 | 0.020 | 0.029 |
| S6 | 0.019 | 0.024 | 0.046 |
| S7 | 0.000 | 0.007 | 0.000 |
| S8 | 0.015 | 0.014 | 0.124 |
| S9 | 0.004 | 0.006 | 0.005 |
| S10 | 19.571 | 24.401 | 30.636 |

## 6. Conclusions

In this paper we produce comprehensive investigation of the feature spaces applicability and effectiveness of different classifiers in solution of the spoofing detection task for ASVspoof Challenge 2015. The submitted STC systems were based on TV modelling in the space of different features and a linear SVM-based classifier or a nonlinear DBN-based classifier.

During our research we tested different front-end features, including features derived from phase spectrum and features obtained by applying wavelet-transform, in order to determine reliable and robust countermeasures in the anti-spoofing task. Experiments performed on the datasets of the Challenge demonstrate that using phase-based and wavelet-based features provides a substantial input into the efficiency of the resulting STC systems. Our best result on the evaluation dataset is 1.965% EER for all spoofing attacks.